# Determination of Correction Factors for Small Field Based on Measurement and Numerical Calculation using Cylindrical Ionization Chambers

**Kwangwoo Park (first author)**
*Department of Radiation Oncology, College of Medicine, Yonsei University, Seoul 120-740*
**Wonhoon Choi (second author)**
*Department of Radiation Oncology, College of Medicine, Yonsei University, Seoul 120-740*
**Sungho Park (third author)**
*Department of Neuro Surgery, Ulsan University Hospital, Ulsan 682-714*
**Jin Hwa Choi (forth author)**
*Department of Radiation Oncology, College of Medicine, Chung-Ang University, Seoul 156-755*
**Suk Won Park (fifth author)**
*Department of Radiation Oncology, College of Medicine, Chung-Ang University, Seoul 156-755*
**Jino Bak (corresponding author)**
*Department of Radiation Oncology, College of Medicine, Chung-Ang University, Seoul 156-755*

We studied the investigation of volume averaging effect for air-filled cylindrical ionization chambers to determine correction factors in small photon field for the given chamber. As a method, we measured output factors using several cylindrical ionization chambers and calculated with mathematical method similar to deconvolution in which we modeled non-constant and inhomogeneous exposure function in the cavity of chamber. The parameters in exposure function and correction factors were determined by solving a system of equations we developed with measurement data and geometry of the given chamber. Correction factors (CFs) we had found are very similar to that from Monte Carlo (MC) simulation. For example, CFs in this study were computed as 1.116 for PTW31010, and 1.0225 for PTW31016, while CFs from MC were reported as approximately between 1.17 and 1.20 for PTW31010, and between 1.02 and 1.06 for PTW31016 in $1 \times 1 cm^2$ **of** 6MV photon beam . Furthermore, the result from the method of deconvolution combined with MC for chamber response function, also showed similar CF for PTW 30013, which was reported as 2.29 and 1.54 in $1\times 1 cm^2$ and $1.5\times 1.5 cm^2$ filed size respectively. The CFs from our method provided similarly as 2.42 and 1.54. In addition, we reported CFs for PTW30013, PTW31010, PTW31016, IBA FC23-C, and IBA CC13. As a consequence, we suggested a newly developed method to measure correct output factor using the fact that inhomogeneous exposure, force a volume averaging effect in a cavity of air filled cylindrical ionization chamber. The result from this method is very similar to that from MC simulation. The method we developed can easily be applied to clinic.

Email: njanka@hanmail.net
Fax: +82-2-2019-3144

# I. INTRODUCTION

The use of small fields in radiotherapy techniques has been increased in recent decade. Especially, such small fields are applied to intensity modulated radiation therapy(IMRT), stereotactic treatments, the frequency of which become growing in modern radiation therapy.

However, it was reported that the results of measurements using several cylindrical ionization chambers were different to each other[1] and in the calibration of output factor based on the protocol recommended by TRS 398[2]. The reason why these differences occurred in measurement, has been studied in many literatures, where they explained that volume averaging effect and breaking of charged particle equilibrium (CPE) in the sensitive volume of detector caused problems of both the discrepancy and under dose measurement in small field size [3-5].

Much efforts for solving these problems have been made, which could be categorized into two methods: deconvolution and Monte Carlo (MC) simulation. The first is that using chamber response function, correct measurement can be extracted as follows [6-10].

$$D_m(x) = \int D_p(u) K(u-x) du \tag{1}$$

where correct measurement, $D_p(u)$ can be obtained from the inverse of the kernel, or chamber response function $K(u-x)$. The latter is that correction factors (CFs) for small field and non-reference field, $k_{Q_{clin},Q_{msr}}^{f_{clin},f_{msr}}$ were introduced and calculated using Monte Carlo simulation [10-14]. However, the modeling of MC is not easy to be used in every clinic.

In this study, we measured output factors using five cylindrical ionization chambers and calculated with mathematical method similar to deconvolution in which we modeled exposure function in the cavity of chamber. Therefore, with both measured data and the method we developed, we obtained CFs in small photon field for each used chambers.

## II. EXPERIMENTS AND DISCUSSION

**A. Correction factor for given chamber$_i$ and field size $\Gamma \times \Gamma cm^2$ : $CF_i^\Gamma$**

It is well known that absorbed dose from the measurement of ionization chamber can be acquired from Bragg-Gray cavity theory[15], where absorbed dose in medium of interest, $D_{med}$ can be calculated from dose in air, $D_{air}$ as follows [10]:

$$D_{med} = D_{air} \cdot s_{med,air} \cdot \prod_i p_i ,$$

(2)

where $s_{med,air}$ is stopping power ratio from air to medium, $p_i$ is perturbation factors. However, it has also been reported that stopping power ratio in small field varies less than 1% [16]. Therefore, absorbed dose in air, $D_{air}$ which directly connected to collective charge from electrometer when we use ionization chamber, should have a key to determine CF.

In order to consider proper collective charge by air-filled cylindrical ionization chamber, exposure ($X$) plays a crucial role because it is defined as the total charge per unit mass liberated in air, $X = \frac{dQ}{dm}$. When ionized charge produced by X-ray can be entirely collected inside cavity of the chamber under the condition of CPE, the collected charge $Q$ is,

$$Q = \int dQ = \int X dm = \int \rho X dV$$

(3)

where $\rho$ is an air density, and $V$ is a sensitive volume of chamber cavity which can be obtained by the geometry of a cavity, especially, radius, $r$ and length $L$ in the case of cylindrical shape (radius and length for used chambers are summarized in Table 2). If the field size is large enough for exposure to be constant in whole chamber cavity, Eq. (3) shows that collected charge is just proportional to the volume of cavity. For reference field size of irradiation in which is relatively flat compared to the size of cavity, exposure can be considered as constant function and encompass whole cavity. In small field size, however, exposure might not be a constant and not encompass the cavity, in the case of which we should correct the measurement.

Non-constant and inhomogeneous exposure would have the shape of Gaussian as described in Fig.1. In modeling of chamber response function, error functions had been used[17]. Accordingly, we modeled one dimensional exposure $X_\Gamma(x;\alpha,\sigma)$ as a composed error functions with small modification as follows for given field size $\Gamma \times \Gamma cm^2$:

$$X_\Gamma(x;\alpha,\sigma) = A_\Gamma \cdot \underbrace{\frac{Erf\left(\frac{x-\alpha}{\sigma}\right) + Erf\left(-\frac{x+\alpha}{\sigma}\right)}{2Erf\left(\frac{-\alpha}{\sigma}\right)}}_{\equiv \text{normalized exposure } \hat{X}_\Gamma(x;\alpha,\sigma)}$$

(4)

with constant $A_\Gamma$ that can be both maximum value of exposure and correct measurement in the center of cavity. In addition, we introduce parameters $\alpha$ and $\sigma$ which describe field size and penumbra of exposure curve respectively. In Eq. (4), normalized one dimensional exposure, $\hat{X}_\Gamma(x;\alpha,\sigma)$ is defined as the sum of two error functions. For the reminder, error function is defined as follows:

$$Erf(x) = \frac{2}{\sqrt{\pi}} \int_0^x e^{-t^2} dt$$

(5)

In small field size, inhomogeneity of exposure forces the measurement to be averaged over whole cavity volume so that actual measurement would be less than correct value, $A_\Gamma$. While at reference field size, $10 \times 10 cm^2$, exposure should be so constant that there is no volume averaging effect, thereby the measurement over whole volume of cavity is identical to the value of point we interest. Thus, at reference field in which exposure would be flat compared to the cavity size of a chamber, $\hat{X}_\Gamma(x;\alpha,\sigma)$ is constant and normalized to 1.

$$\int_{-\lambda}^{\lambda} \hat{X}_{10}(x;\alpha,\sigma)dx = \int_{-\lambda}^{\lambda} 1 \cdot dx = 2 \cdot \lambda$$

(6)

Now, 1-dimensional exposure can be easily expanded to 2-dimensional function under the assumption of $xy$ symmetry as follows:

$$X_\Gamma(x,y;\alpha,\sigma) = A_\Gamma \cdot \hat{X}_\Gamma(x;\alpha,\sigma) \cdot \hat{X}_\Gamma(y;\alpha,\sigma)$$

(7)

In small cavity volume, because we take approximation that exposure is the same over z direction, three dimensional exposure function can be modeled as Eq. (7). Consequently, collective charge can be derived from Eq. (3) as follows:

$$\begin{aligned}Q &= \rho A_\Gamma C_z \cdot \int_{-r}^{r} \hat{X}_\Gamma(x;\alpha,\sigma)dx \cdot \int_{-\frac{L}{2}}^{\frac{L}{2}} \hat{X}_\Gamma(y;\alpha,\sigma)dy \\ &= \rho A_\Gamma C_z \eta_\Gamma(\alpha,\sigma,r)\eta_\Gamma(\alpha,\sigma,L/2)\end{aligned}$$

(8)

where $C_z$ is a constant from the integration in $z$-component, which does not depend on field size but chamber geometry. Moreover, $C_z$ should be cancelled out when we calculate output factor. For the simplicity, we defined integrated exposure function $\eta_\Gamma(\alpha,\sigma,\lambda)$ as follows,

$$\begin{aligned}\eta_\Gamma(\alpha,\sigma,\lambda) &= \int_{-\lambda}^{\lambda} \hat{X}_\Gamma(x;\alpha,\sigma)dx \\ &= \frac{1}{Erf\left(\frac{\alpha}{\sigma}\right)}\left[\frac{\sigma}{\sqrt{\pi}}\left(e^{-(\alpha+\lambda)^2/\sigma^2} - e^{-(\alpha-\lambda)^2/\sigma^2}\right) + (\alpha+\lambda)Erf\left(\frac{\alpha+\lambda}{\sigma}\right) - (\alpha-\lambda)Erf\left(\frac{\alpha-\lambda}{\sigma}\right)\right]\end{aligned}$$

(9)

Note that for reference field size $10\times 10 cm^2$, $\eta_{10}(\alpha,\sigma,\lambda)$ should satisfy the identity Eq. (6). Therefore, output factor measured by a chamber$_1$ is,

$$OF^{measured}_{chm_1,\Gamma} = \frac{A_\Gamma \eta_\Gamma(\alpha,\sigma,r_1)\eta_\Gamma(\alpha,\sigma,L_1/2)}{A_{10}\eta_{10}(\alpha,\sigma,r_1)\eta_{10}(\alpha,\sigma,L_1/2)}$$

(10)

Similarly, for another chamber$_i$,

$$OF^{measured}_{chm_i,\Gamma} = \frac{A_\Gamma \eta_\Gamma(\alpha,\sigma,r_i)\eta_\Gamma(\alpha,\sigma,L_i/2)}{A_{10}\eta_{10}(\alpha,\sigma,r_i)\eta_{10}(\alpha,\sigma,L_i/2)} = \frac{A_\Gamma \eta_\Gamma(\alpha,\sigma,r_i)\eta_\Gamma(\alpha,\sigma,L_i/2)}{A_{10}\left(\frac{r_i L_i}{r_1 L_1}\right)\eta_{10}(\alpha,\sigma,r_1)\eta_{10}(\alpha,\sigma,L_1/2)}$$

(11)

where we applied

$$\eta_{10}(\alpha,\sigma,r_i) = \int_{-r_i}^{r_i}\hat{X}_{10}(x;\alpha,\sigma)dx = 2r_i = \frac{r_i}{r_1}\cdot \eta_{10}(\alpha,\sigma,r_1)$$

(12)

For determination of $\alpha$ and $\sigma$, we considered the following ratios

$$R^{1,i}_\Gamma = \frac{OF^{measured}_{chm_1,\Gamma}}{OF^{measured}_{chm_i,\Gamma}} = \frac{r_1 L_1 \eta_\Gamma(\alpha,\sigma,r_1)\eta_\Gamma(\alpha,\sigma,L_1/2)}{r_i L_i \eta_\Gamma(\alpha,\sigma,r_i)\eta_\Gamma(\alpha,\sigma,L_i/2)}$$

(13)

where $i = $ 2 and 3 denote three used chambers whose volumes should be different. A system of equations consisting of $R^{1,2}_\Gamma$ and $R^{1,3}_\Gamma$ can be exactly solved for unkowns $\alpha$ and $\sigma$. In Table 1, solutions for the system of equations summarized. Equipped with $\alpha$ and $\sigma$, CF can be derived from Eq. (10).

$$OF_{chm_i,\Gamma}^{measured} = \frac{A_\Gamma}{A_{10}} \cdot \frac{\eta_\Gamma(\alpha,\sigma,r_i)\eta_\Gamma(\alpha,\sigma,L_i/2)}{\eta_{10}(\alpha,\sigma,r_i)\eta_{10}(\alpha,\sigma,L_i/2)} = OF_{chm_i,\Gamma}^{corredcted} \cdot \frac{\eta_\Gamma(\alpha,\sigma,r_i)\eta_\Gamma(\alpha,\sigma,L_i/2)}{2 \cdot r_i \cdot L_i}$$

(14)

Finally we define inhomogeneous exposure CF for given field size, $\Gamma$ and chamber $_i$, $CF_i^\Gamma$ as follows:

$$CF_i^\Gamma = \frac{2 \cdot r_i \cdot L_i}{\eta_\Gamma(\alpha,\sigma,r_i) \cdot \eta_\Gamma(\alpha,\sigma,L_i/2)}$$

(15)

Corrected output factor is, then, simply acquired.

$$OF_{chm_i,\Gamma}^{corredcted} = OF_{chm_i,\Gamma}^{measured} \cdot CF_i^\Gamma$$

(16)

**B. Output factors measurement**

Output factor measurement was performed using linear accelerator, Elekta Infinity™ with multi-leaf collimator, MLCi2. Prior to actual measurment, we adjust a center of radiation field using lateral and longitudinal profile curves of $20 \times 20 cm^2$ at depths of $5cm$ and $10cm$ with source to surface distance (SSD), $90cm$. All output factors were measured at $90cm$ SSD and $10cm$ depth, in which temperature and pressure correction are also applied. Measurement data are plotted in Fig.2 (6MV) and Fig.3 (10MV).

**C. Results and Discussion**

We numerically solved a system of equations, Eq. (13) combined with output factors measured by chamber 1, 2, and 3. PTW30013 was fixed as chamber 1, while for chamber 2 and 3, six possible combinations of four chambers, PTW31010, PTW31016, IBA FC23-C, and IBA CC13 used for the determination of $\alpha$ and $\sigma$. As an example, the combination of CC13(Chm2) and PTW31016(Chm3) provided output factor ratios in $1 \times 1 cm^2$ field size of 6MV photon beam, $R_{1\times1}^{1,2} = 0.46555$ and $R_{1\times1}^{1,3} = 0.42985$ which were calculated from the measured output factors, $OF_{chm_1,1\times1}^{measured} = 0.2677$, $OF_{chm_2,1\times1}^{measured} = 0.5750$, and $OF_{chm_3,1\times1}^{measured} = 0.6227$. Another set of equations was acquired by integration $\eta_{1\times1}(\alpha,\sigma,\lambda)$ of Eq. (13),

$$0.46555 = \frac{3.05 \cdot 23 \cdot \eta_{1\times1}(\alpha,\sigma,3.05)\eta_{1\times1}(\alpha,\sigma,23/2)}{3 \cdot 5.8 \cdot \eta_{1\times1}(\alpha,\sigma,3)\eta_{1\times1}(\alpha,\sigma,5.8/2)}$$

(17)

$$0.42985 = \frac{3.05 \cdot 23 \cdot \eta_{1\times1}(\alpha,\sigma,3.05)\eta_{1\times1}(\alpha,\sigma,23/2)}{1.45 \cdot 2.9 \cdot \eta_{1\times1}(\alpha,\sigma,1.45)\eta_{1\times1}(\alpha,\sigma,2.9/2)}$$

(18)

Solution of above system of equations was numerically computed as $\alpha = 4.986mm$ and $\sigma = 3.129mm$. In Table 1, we summarized average and standard deviation of $\alpha$ and $\sigma$ for the case of 6MV and 10MV photon beams. In larger field size than $3.5\times3.5cm^2$, integrated exposure function, $\eta_{\Gamma}(\alpha,\sigma,\lambda)$ is too slowly varying to solve a system of equation. However, percentage differences between output factors measured by various chambers are at most $0.8\%$, where CFs caused by inhomogeneous exposure are considered as 1. Based on determined $\alpha$ and $\sigma$, we plotted relative exposure functions of 10MV photon for various small fields in Fig.1(b) where we verified exposure could not encompass entire cavity of chamber in small fields. From solutions of $\alpha$ and $\sigma$, we calculated CFs based on Eq.(15) which are tabulated in Table 2. Corrected output factors from six combinations of four different chambers showed small standard deviations, i.e., within $3\%$. Therefore, there is little chamber dependency based on the method in this study.

In Fig.2 and Fig. 3., we showed how well output factors are corrected after CFs were applied, where all output factors from various chambers were converged compared to non-corrected measured output factors, which were showing the property of chamber dependency. In addition, output factors measured by PTW60016 (diode detector) was shown for easy comparison. Moreover, for small field dosimetry, diode detector is usually recommended owing to tiny sensitive volume. However detector response of diode can vary depending on fluence especially in small field size, where diode measurement might exceed actual value [18, 19]. CF for PTW60016 and other diode detectors were reported as about $0.95$ in many literatures [12-14, 18, 20, 21]. Fig. 4 shows CFs of various chambers versus field sizes from $1\times 1 cm^2$ to $3.5\times 3.5 cm^2$, where dashed curves were fitted using exponential function, $ae^{-bx}+1$ with fitting parameters $a$ and $b$. CFs for inhomogeneous exposure, thereby fluence, are comparable to that for small and nonstandard fields suggested by Alfonso *at. el.* [11], $k_{Q_{clin},Q_{msr}}^{f_{clin},f_{msr}}$, because $CF_i^{\Gamma}$ from output factor can be applied to only a measurement in small field size, which is the numerator of output factor in Eq.(10). Note that reference field size which is the denominator of output factor in Eq.(10) does not need inhomogeneous exposure correction. Correction factor, $k_{Q_{clin},Q_{msr}}^{f_{clin},f_{msr}}$ for various chambers had been reported that Monte Carlo simulated $k_{Q_{clin},Q_{msr}}^{f_{clin},f_{msr}}$ for 6MV photon beam (Siemens KD linear accelerator) was approximately between $1.17$ and $1.22$ for PTW31010, and between $1.02$ and $1.06$ for PTW31016 [18]. CFs in this study were computed as $1.143$ for PTW31010, and $1.024$ for PTW31016. Even though we used different linear accelerator, those two values which are computed by MC and our method are very similar. Furthermore, the result from the method of deconvolution combined with MC for chamber response function, also showed similar CF for PTW 30013, which was found as 2.29 and 1.54 in 1x1 and 1.5x1.5 cm2 filed size respectively [10]. The CFs from our method provided similarly as 2.42 and 1.54. In addition, we reported CFs for PTW30013, PTW31010, PTW31016, IBA FC23-C, and IBA CC13.

We also computed $\alpha$ and $\sigma$ for all possible combinations of five chambers, which provided very small standard deviation. So, we could verify the property of chamber independency to determine exposure function, i.e., $\alpha$ and $\sigma$.

### III. CONCLUSION

In this study, we suggested a newly developed method to measure correct output factor using the fact that in-homogeneous exposure which is directly connected with photon fluence, force a volume averaging effect in a cavity of air filled cylindrical ionization chamber. The result from this method is very similar to that of Monte Carlo simulation. Furthermore, this method can easily extend to solve any other problems occurring in small photon field size. However, problems in the region of breaking CPE are still under research topic.

**ACKNOWLEGEMENTS**

This work was supported by the Nuclear Safety Research Program through the Korea Radiation Safety Foundation(KORSAFe) and the Nuclear Safety and Security Commission(NSSC), Republic of Korea (Grant No. 1305033)

**REREFENCES**

Table 1. Solutions of parameters in exposure function of Eq.(4)

| Energy | parameter | Field size | 1 | 1.5 | 2 | 2.5 | 3 |
|---|---|---|---|---|---|---|---|
| 6MV | $\alpha$ | average | 4.7979 | 7.5885 | 9.9601 | 16.7019 | 20.2363 |
| | | std | 0.1828 | 0.0076 | 0.0205 | 0.2838 | 0.1542 |
| | $\sigma$ | average | 3.4656 | 3.6258 | 3.7755 | 10.2512 | 9.3925 |
| | | std | 0.2820 | 0.0897 | 0.0689 | 0.5558 | 0.1406 |
| 10MV | $\alpha$ | average | 5.1295 | 7.7148 | 10.1097 | 14.2401 | 18.6847 |
| | | std | 0.0083 | 0.0031 | 0.2487 | 0.0160 | 0.6195 |
| | $\sigma$ | average | 3.3665 | 3.6693 | 5.6287 | 9.5239 | 10.2109 |
| | | std | 0.0161 | 0.0470 | 2.5361 | 0.0591 | 0.7882 |

Table 2. Results of corrected output factors and correction factors for 6MV and 10 MV

| Energy | Chamber | Field size | | | | |
|---|---|---|---|---|---|---|
| | | 1 | 1.5 | 2 | 2.5 | 3 |
| 6MV | PTW 30013 | 2.382(3) | 1.540(3) | 1.2126(3) | 1.07620(5) | 1.02463(5) |
| | PTW 31010 | 1.116(2) | 1.020(2) | 1.0023(4) | 1.00735(6) | 1.00170(6) |
| | PTW 31016 | 1.0225(1) | 1.0031(2) | 1.00030(2) | 1.00168(4) | 1.00039(0) |
| | FC 23-C | 1.207(1) | 1.040(2) | 1.0053(2) | 1.0121(2) | 1.00288(3) |
| | CC 13 | 1.110(2) | 1.019(2) | 1.0021(4) | 1.00704(5) | 1.00163(6) |
| 10MV | PTW 30013 | 2.294(2) | 1.517(2) | 1.2181(7) | 1.12162(8) | 1.0479(1) |
| | PTW 31010 | 1.112(2) | 1.018(1) | 1.0071(7) | 1.0128(1) | 1.0046(1) |
| | PTW 31016 | 1.02240(1) | 1.0035(1) | 1.00149(5) | 1.002896(1) | 1.00104(5) |
| | FC 23-C | 1.1949(1) | 1.0356(9) | 1.0135(4) | 1.02067(1) | 1.0073(1) |
| | CC 13 | 1.106(2) | 1.017(1) | 1.0068(7) | 1.01226(9) | 1.0044(1) |

Figure Captions.

Fig. 1. (a) Gaussian modeling of exposure (b) Relative exposure functions of 10MV photon for various field size, which were calculated from $\alpha$ and $\sigma$ in Table. 1.

Fig. 2. 6MV output factors (a) without correction (b) with correction

Fig. 3. 10MV output factors (a) without correction (b) with correction

Fig. 4. 10MV(a) and 6MV(b) correction factors